\documentclass[lettersize,journal]{IEEEtran}
\usepackage{amsmath,amsfonts}
\usepackage{algorithmic}
\usepackage{algorithm}
\usepackage{array}
\usepackage[caption=false,font=normalsize,labelfont=sf,textfont=sf]{subfig}
\usepackage{textcomp}
\usepackage{stfloats}
\usepackage{url}
\usepackage{verbatim}
\usepackage{graphicx}
\usepackage{cite}
\usepackage[table]{xcolor}
\usepackage{float}
\def\BibTeX{{\rm B\kern-.05em{\sc i\kern-.025em b}\kern-.08em
		T\kern-.1667em\lower.7ex\hbox{E}\kern-.125emX}}

\begin{document}

\title{ Theoretical  Grid-Forming Extreme of  Inverters }

 \author{Qianxi~Tang,~\IEEEmembership{Graduate Student Member,~IEEE,}
	Li~Peng,~\IEEEmembership{Senior Member,~IEEE}
        % <-this % stops a spaceF
\thanks{% <-this % stops a space
This work has been submitted to the IEEE for possible publication. Copyright may be transferred without notice, after which this version may no longer be accessible.
%The article was created thanks to participation in program PROM of the Polish National Agency for Academic Exchange. The program is co-financed from the European Social Fund within  the Operational Program Knowledge Education Development, non-competitive project entitled ?International scholarship exchange of PhD students and academic staff? executed under the Activity 3.3 specified in the application for funding of project No. POWR.03.03.00-00-PN13/18. The work has been also supported by the Fundamental Research Funds for the Central Universities project no.~21620335.
(Corresponding author: Li~Peng.)

Qianxi~Tang and  Li~Peng are with the State Key Laboratory of Advanced Electromagnetic Engineering and Technology, School of Electrical and Electronic Engineering, Huazhong University of Science and Technology, Wuhan 430074, China (e-mail: qianxi@hust.edu.cn; pe105@mail.hust.edu.cn;).
}
\markboth{Journal of \LaTeX\ Class Files,~Vol.~14, No.~8, August~2021}%
{Shell \MakeLowercase{\textit{et al.}}: A Sample Article Using IEEEtran.cls for IEEE Journals}

\IEEEpubid{0000--0000/00\$00.00~\copyright~2021 IEEE}
}
\maketitle

\begin{abstract}
What are the theoretical and physical limits of a grid-forming inverter? This letter proposes that the extreme grid-forming ability of inverters is limited by their dc-side, ac-side, circuit topology dynamics, but not control. While many papers focus on how to improve grid-forming inverters stability, power sharing, inertia emulation, fault response — few, if any, formally define the fundamental theoretical limits or extremes of grid-forming behavior. It seems that the grid-forming can be improved endlessly. No physical system can support a grid indefinitely without limitations, especially under increasing levels of disturbance or uncertainty. Therefore, this boundary is explicitly shown by a mathematical expression in this letter. Consequently, the results show that relatively low dc-side voltage and high active power injection could damage the grid-forming ability. Poor consideration of dc-side, ac-side, and circuit topology dynamics in real practice will cause jeopardizing oscillation even by the theoretical best grid-forming control strategy.
\end{abstract}


\section{Introduction}
\IEEEPARstart{T}{he}  stability analysis and control designing of a grid-forming inverter (GFMI) is sophisticated due to the complex interactions between its control strategy, dc-side dynamics, ac-side dynamics, and power electronics circuit dynamics \cite{1}. Different control methods, such as matching, virtual synchronous, virtual oscillator,  feedback linearization control .etc, have been proposed to prevent the instability when GFMI encounters ac-side complex voltage disturbance, complex power change, and grid strength changing \cite{2}. However, they neglect the dc-side dynamics which might have a unique dynamics when the dc-link is not so stiff. Although, in some articles, dc-link dynamics has been embedded into the synchronization loop to enhance power or complex voltage reference tracking performance \cite{3,4,5}, the dc-side effect to the voltage control loop has been ignored or simplified and few articles discuss within what extend ac and dc disturbances could be, the grid-forming inverter can still maintain its supportive role to the grid. More enhanced control strategies for grid-forming are emerging  \cite{6,7,10}, but few studies, if any, examine where the theoretical highest performance of a grid-forming can be and what are the factors that limit it.

This letter illustrates that the voltage controllability is a function of the topology, dc-side voltage, and ac-side current (TDA). Namely, the accuracy of tracking the reference voltage or power is limited by these three factors. The result gives out this explicit quantified relationship under approximations. It shows that such an mathematical expression can be used to determine the safe operable area of a grid-forming inverter under small/large dc-link voltage and load change. It finds out that it is the TDA that determines the extreme theoretical tracking limitation of grid-forming inverters. 

The rest of this letter is organized as follows. Section II
defines what is the theoretical extreme grid-forming capability. Section III derives the explicit mathematical expression for the theoretical extreme forming ability of a specific inverter, trying to illustrate the core idea without losing generalization. The experimental verification is in Section IV. Finally, Section V concludes this letter.
\IEEEpubidadjcol 

\section{The Absolute Physical Boundary of a Grid-Forming Inverter’s Capability}
The capability of a grid-forming inverter fundamentally lies in its ability to establish and regulate voltage and frequency within a power system. This core function requires tracking predefined dynamic objectives despite grid disturbances and parameter uncertainties. Whether the inverter is maintaining synchronization, regulating power flow, or enabling fault ride-through, each function is essentially a manifestation of robust trajectory tracking. A broad class of advanced control strategies—ranging from damping-enhancement schemes \cite{8} to impedance-robust synthesis and dynamic loop shaping—are not aimed at redefining this role, but rather at expanding the conditions under which reliable tracking is possible. Once tracking is lost, the inverter can no longer fulfill its grid-forming purpose. This consistent reliance on tracking behavior underlines a critical observation: grid-forming capability is inherently bounded by the inverter’s tracking ability. As these designs evolve, what varies is the margin of this tracking boundary, not the nature of the boundary itself.

Let us assume that the optimal reference states for grid-forming inverters at the point of common coupling (PCC) have been established through an ideal synchronization mechanism. These reference states, comprising voltage magnitude, phase, and frequency, can therefore be equivalently represented as a time-varying voltage reference signal. Under this assumption, if the inverter’s terminal voltage precisely tracks the reference signal, then by definition, it achieves the ideal synchronization state. Hence, good voltage tracking guarantees good synchronization performance. Conversely, if voltage tracking is poor, meaning there is a persistent error in magnitude, frequency, or phase, then the inverter's terminal voltage necessarily deviates from the reference synchronization state. Therefore, poor tracking guarantees poor synchronization. By the contrapositive of this implication, good synchronization performance cannot exist without good voltage tracking. Together, these two directions establish a bidirectional condition: Good voltage tracking is both necessary and sufficient for good synchronization or grid-forming performance, provided the ideal reference is known and accurately defined.

The above demonstrates that, under the assumption of an ideal and accurate synchronization reference, good voltage tracking is both a necessary and sufficient condition for good synchronization performance. This establishes the inverter’s voltage tracking limit as the theoretical boundary of its grid-forming capability. While practical considerations, such as reference accuracy, internal constraints, and dynamic transients, may influence operational behavior, the proposed condition provides a fundamental benchmark for evaluating and comparing grid-forming strategies.
\begin{figure}[h]
	\centering
	\includegraphics[width=3in]{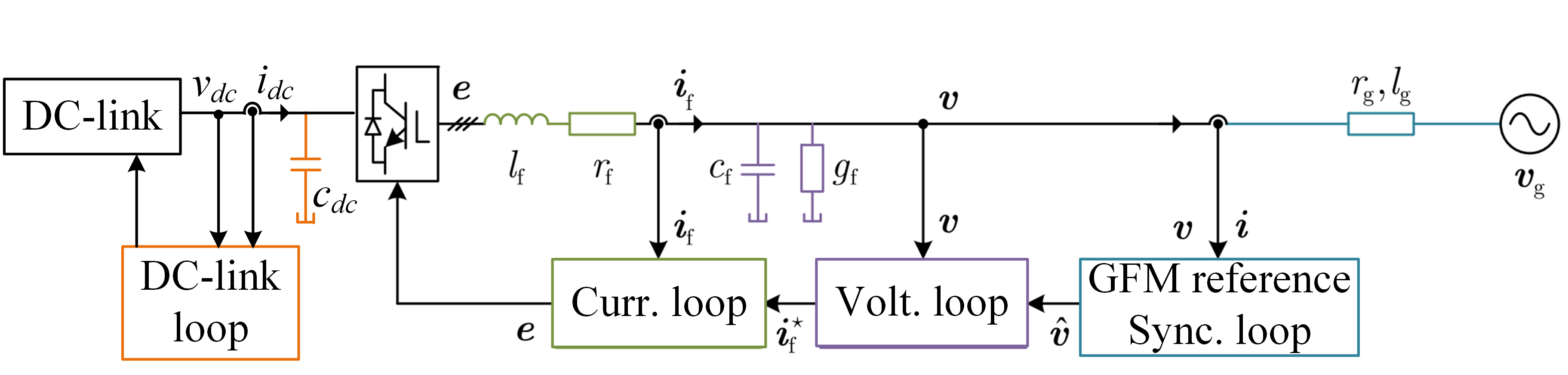}
	\caption{The GFMI system.}
	\label{fig23}
\end{figure}

\section{ Physical Voltage Tracking Boundary of a Grid-Forming Inverter}
For simplifying the derivation of voltage tracking limit, the one phase H-bridge inverter is chosen. However, the key idea of getting the physical boundary and the method below can be extended to other circuit topologies. Therefore, this simplification should not be considered as a loss of generality.

Fig.~\ref{fig23} illustrates the GFMI system with a dc-link, an output LC filter and a transmission line. According to [Author, Year], the state-space model is given by \eqref{eq3}. The variables are capacitor voltage $u_o$, output current $i_o$, switching variable $T$, and dc-link $v_{dc}$ measured by sensors.
\begin{equation}
	\begin{aligned}
		\label{eq3}
		&\frac{d^2 u_o}{d^2 t}=-\frac{u_o}{L_f C_f}-\frac{1}{C_f} \frac{d i_o}{d t}+\frac{T v_{dc}}{L_f C_f} \\
		&\frac{d^2 u_o}{d^2 t}=-\frac{u_o}{L_f C_f}-\frac{1}{C_f} \frac{d i_o}{d t}+u=f\left(u_o, i_o\right)+u \\
	\end{aligned}
\end{equation}

To simplify the analysis and reduce the system's order, the second-order dynamic equation \eqref{eq3} is replaced by a first-order approximation involving a new positive constant~$\lambda$ and newly defined variables $\tilde{x}$ and $s$, as shown in~\eqref{eq4}, where $u_o$ is $x$ and $x_d$ is the desired reference. So, the solution of \eqref{eq4} is shown by the \eqref{eq45}. 
\begin{equation}
	\begin{aligned}
			\label{eq4}
		&\tilde{x}=x-x_d\\
		&s(x, t)=\left(\frac{d}{d t}+\lambda\right) \tilde{x}
		\end{aligned}
\end{equation}

\begin{equation}
	\label{eq45}
	\tilde{x}(t) = e^{-\lambda t}\left[\tilde{x}(t_0)e^{\lambda t_0} + \int_{t_0}^{t} s(x,\tau)e^{\lambda \tau} d\tau\right]
	\end{equation}

Given any function $s(x,t)$ and the initial conditions at $t_0$ in \eqref{eq45} , with integrable $s(x,t)$  in this system, the existence of at least one $\tilde{x}$ fulfilling the equation guarantees surjectivity. The injective is easy to prove. Therefore, with specified initial conditions, the linear operator is indeed bijective. So, $s(x,t)$ and $\tilde{x}$ are equaivalent with the known initial conditions which are measured by sensors. Choosing $\lambda= \frac{f_{\mathrm{sw}}}{10}$ where $f_{sw}$ is the designed nominal switching frequency. Consequently, with the traditional design method of LC filter, $f_{\mathrm{sw}} = a\cdot \frac{1}{2\pi \sqrt{LC}}, a\in [10, \infty)$. Accurate tracking the reference means keeping $\tilde{x}$ zero. To control the  $s(x,\tau)$, \eqref{eq6sdot} is obtained and the $u$ term shows that $s(x,\tau)$ is a sawtooth-like waveform where the order of magnitude of $\ddot{x}_d$ and $\lambda \dot{\tilde{x}}$ is relatively small and can be ignored. 
\begin{equation}
	\label{eq6sdot}
	\dot{s}=f-\ddot{x}_d+\lambda \dot{\tilde{x}}+u.
	\end{equation}
	
So, to ensure controllability, the condition given in \eqref{eq56} must be satisfied.
\begin{equation}
	\label{eq56}	
	\frac{v_{dc}}{L_f C_f}>\left| \frac{x}{L_f C_f} + \frac{1}{C_f} \frac{d i_o}{d t} \right| 
	\end{equation}

 This result shows that relatively low dc-side voltage, high output voltage, high active power injection could damage the absolute physical boundary of a grid-forming inverter’s capability. It is noted that a sawtooth-like waveform of $s(x,\tau)$ cannot keep $\tilde{x}$ zero and it must have an error there. Now, below \eqref{eq67} is the quantification of this effect, where $ s(t_0) = k \frac{v_{dc}}{L_f C_f} \Delta t$, with the empirically chosen constant $ k\in (0, 1) $. The larger the left-hand side of \eqref{eq56}, the closer the value of $ k$  is to 1. $t_0$ is the time whenever the  $s(x, t)$ is larger than $\text{Bound}$ and $J$ is defined as 1 when $s(x, t)>\text{Bound} $, else 0.
\begin{align}
	\label{eq67}
	\tilde{x}(t) &\approx \left[ u(t) - u(t - \Delta t - t_0) \right] 
	(t - t_0)  (-\frac{k v_{\mathrm{dc}}}{L_f C_f \lambda} )J \nonumber \\
	&\quad + \left[ u(t) - u(t - \Delta t - t_0) \right] x(t_0)J \nonumber \\
	&\quad + \frac{\mathrm{Bound}}{\lambda} \sin(2\pi f_{\mathrm{sw}} t)\, u(t - \Delta t - t_0), \nonumber \\
	&\quad t \in [t_0, \infty), \quad \mathrm{Bound} \in [H_b, \infty)
\end{align}

 Equation \eqref{eq67} shows that the best tracking performance is related to the level of $v_{\mathrm{dc}}$ and the Bound term, $|s(t)|\leq \text{Bound}$, which is a  parameter. If the Bound term is smaller, the tracking would be better. However, the accuracy of calculation  $s(t)$ is limited to  sampling and delay caused by the hardware. Meanwhile, the smaller Bound assigned, the higher  switching frequency could be. Therefore, the parameter Bound admits a minimum value, denoted
 $H_b(f_{\mathrm{sw}}, \mathrm{sampling}, \mathrm{delay} )$. Thus far,  \eqref{eq67}, the voltage tracking boundary of the inverter,  has been formally derived. It should be noted that the configuration of all parameters in this section is for mathematical transformation and analysis not for designing specific control strategy. 
 \begin{figure*}
\centering
\includegraphics{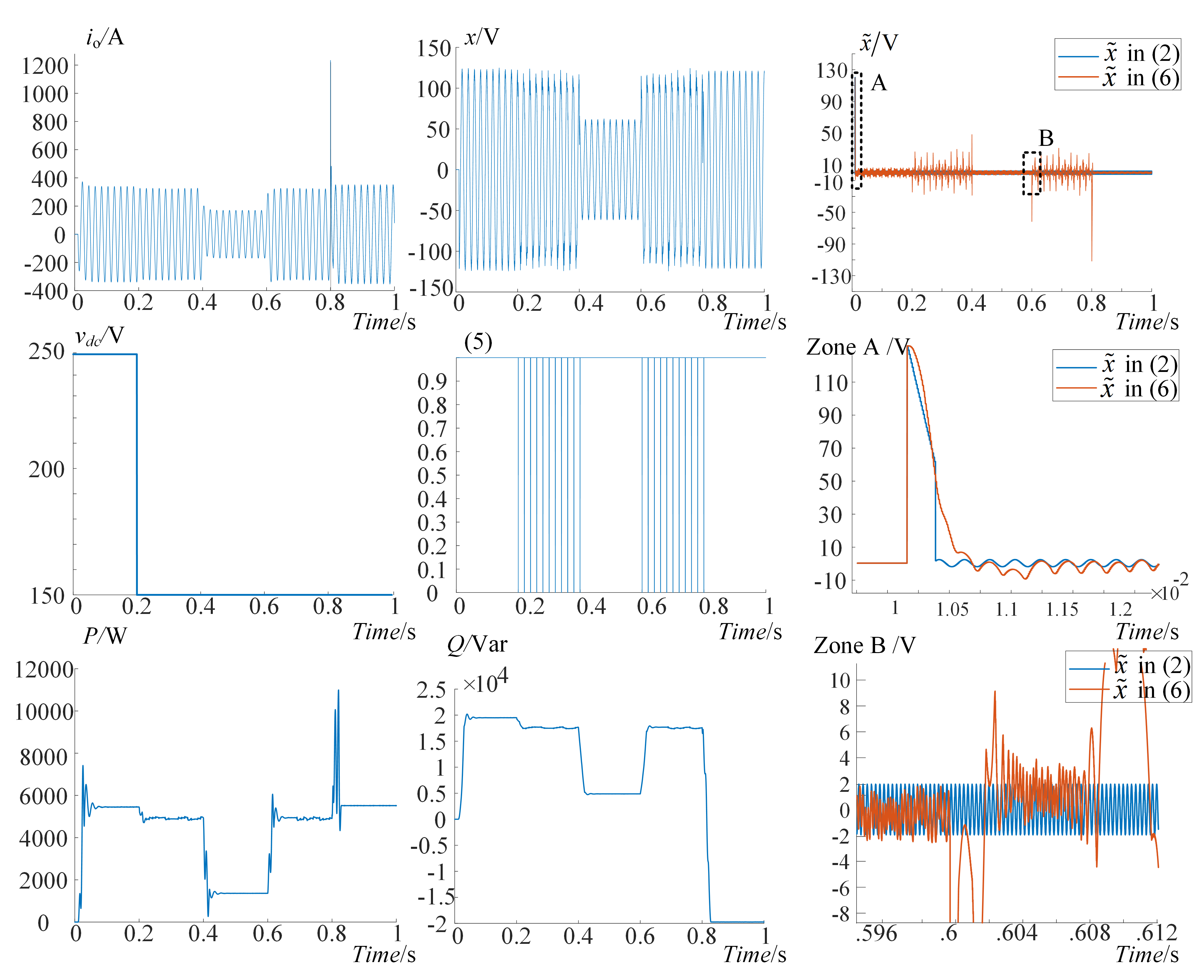}%
\caption{The boundary condition simulation test.
\label{fig_first_case}}
\end{figure*}

\section{Verification of the Best Tracking Performance}
Based on the best tracking notion in section III, an enhanced symmetric sliding mode control \cite{9} is used in one phase GFMI system to show the error between this practical control and the theoretical best tracking strategy performance. Meanwhile, the operation points are dangerous because the current is large, so the verification is done by simulation.
The influence of the variables \( v_{\mathrm{dc}} \), \( \frac{d i_o}{d t} \), and \( x \) on the voltage tracking capability is investigated to evaluate the validity of the physical boundary condition expressed in~\eqref{eq56}. While \eqref{eq67} serves as an approximate relation, the condition defined by~\eqref{eq56} provides a more accurate assessment of the inverter's controllability limits.

A test scenario is considered in Fig.~\ref{fig_first_case} which the dc-link voltage \( v_{\mathrm{dc}} \) is reduced from 250\,V to 150\,V at \( t = 0.2\,\mathrm{s} \), under a load profile of 5\,kW active power and 20\,kVar inductive reactive power. According to the criterion in~\eqref{eq56}, this operating condition exceeds the physical tracking boundary, resulting in a loss of normal voltage tracking performance.

To restore compliance with the criterion, the reference signal \( x_d \) is reduced to 60\,V at \( t = 0.4\,\mathrm{s} \). Under this condition, the inequality in~\eqref{eq56} is satisfied, and the voltage tracking behavior returns to normal. Subsequently, at \( t = 0.6\,\mathrm{s} \), the reference \( x_d \) is increased back to 120\,V, again violating the condition and resulting in degraded tracking performance.

Finally, the 20\,kVar inductive load is replaced with a 20\,kVar capacitive at \( t = 0.8\,\mathrm{s} \) load while maintaining the same voltage reference. This modification brings the system back within the valid tracking region defined by~\eqref{eq56}, restoring satisfactory voltage tracking behavior.

These observations confirm that the inequality in~\eqref{eq56} provides an effective and practical criterion for assessing the feasibility of voltage tracking under various loading and voltage conditions.

\section{Conclusion}
This paper focused on the physical grid-forming boundary of inverters. It finds out that the grid-forming capability is the voltage controllability after assuming the optimal voltage reference has been got. High PCC voltage, low dc-link voltage and fast change of output alternating current might damage the voltage tracking controllability and we have got the explicit expression of this boundary and the approximate best tracking performance. The ability to maintain grid-forming is no longer solely determined by the control strategy, but is also critically influenced by the surrounding operating environment. For example, when the output current of an GFMI varies really fast because of a load  disturbance or grid strength change, the voltage controllability might be lost, which could cause fault ride-through strategy involved, like current limiting. Therefore, a careful examination of controllability criterion under all possible operating conditions for inverters is essential to ensure reliable voltage tracking and overall grid-forming capability.

\bibliographystyle{IEEEtran}
\bibliography{IEEEabrv,RS1}

\begin{thebibliography}{10}
\providecommand{\url}[1]{#1}
\csname url@samestyle\endcsname
\providecommand{\newblock}{\relax}
\providecommand{\bibinfo}[2]{#2}
\providecommand{\BIBentrySTDinterwordspacing}{\spaceskip=0pt\relax}
\providecommand{\BIBentryALTinterwordstretchfactor}{4}
\providecommand{\BIBentryALTinterwordspacing}{\spaceskip=\fontdimen2\font plus
\BIBentryALTinterwordstretchfactor\fontdimen3\font minus
  \fontdimen4\font\relax}
\providecommand{\BIBforeignlanguage}[2]{{%
\expandafter\ifx\csname l@#1\endcsname\relax
\typeout{** WARNING: IEEEtran.bst: No hyphenation pattern has been}%
\typeout{** loaded for the language `#1'. Using the pattern for}%
\typeout{** the default language instead.}%
\else
\language=\csname l@#1\endcsname
\fi
#2}}
\providecommand{\BIBdecl}{\relax}
\BIBdecl

\bibitem{1}
J.~Huang, X.~Yuan, and S.~Wang, ``Power-imbalance stimulation and
  internal-voltage response relationships based modeling method of
  pe-interfaced devices in dc voltage control timescale,'' \emph{IEEE Access},
  vol.~11, pp. 105\,214--105\,224, 2023.

\bibitem{2}
W.~Si, J.~Fang, X.~Chen, T.~Xu, and S.~M. Goetz, ``Transient angle and voltage
  stability of grid-forming converters with typical reactive power control
  schemes,'' \emph{IEEE Journal of Emerging and Selected Topics in Power
  Electronics}, pp. 1--1, 2024.

\bibitem{3}
R.~Mourouvin, T.~Nurminen, M.~Hinkkanen, and M.~Routimo, ``Direct dc-bus
  control for grid-forming converters: Toward the concept of
  dual-voltage-forming converters,'' \emph{IEEE Transactions on Power
  Electronics}, vol.~40, no.~6, pp. 7789--7799, 2025.

\bibitem{4}
C.~Xu, Z.~Zou, X.~Liu, M.~Huang, W.~Chen, and Z.~Wang, ``Stability analysis and
  control design of grid-forming converters with dc-link effect,'' \emph{IEEE
  Transactions on Power Electronics}, vol.~40, no.~5, pp. 6813--6828, 2025.

\bibitem{5}
C.~Luo, X.~Ma, T.~Liu, and X.~Wang, ``Adaptive-output-voltage-regulation-based
  solution for the dc-link undervoltage of grid- forming inverters,''
  \emph{IEEE Transactions on Power Electronics}, vol.~38, no.~10, pp.
  12\,559--12\,569, 2023.

\bibitem{6}
C.~Ai, Y.~Li, Z.~Zhao, Y.~Gu, and J.~Liu, ``An extension of grid-forming: A
  frequency-following voltage-forming inverter,'' \emph{IEEE Transactions on
  Power Electronics}, vol.~39, no.~10, pp. 12\,118--12\,123, 2024.

\bibitem{7}
S.~Chakraborty, S.~Patel, and M.~V. Salapaka, ``-synthesis-based generalized
  robust framework for grid-following and grid-forming inverters,'' \emph{IEEE
  Transactions on Power Electronics}, vol.~38, no.~3, pp. 3163--3179, 2023.

\bibitem{10}
F.~Milano, ``Dual grid-forming converter,'' \emph{IEEE Transactions on Power
  Systems}, vol.~40, no.~2, pp. 1993--1996, 2025.

\bibitem{8}
A.~Tayyebi, D.~Groß, A.~Anta, F.~Kupzog, and F.~Dörfler, ``Frequency
  stability of synchronous machines and grid-forming power converters,''
  \emph{IEEE Journal of Emerging and Selected Topics in Power Electronics},
  vol.~8, no.~2, pp. 1004--1018, 2020.

\bibitem{9}
Q.~Tang and L.~Peng, ``A slack bus grid-forming inverter based on symmetric
  sliding mode control against power sharing imbalances among
  microgenerators,'' in \emph{IECON 2024 - 50th Annual Conference of the IEEE
  Industrial Electronics Society}, 2024, pp. 1--6.

\end{thebibliography}

\end{document}